%% file: ms.tex
\documentclass{emulateapj}
\usepackage{natbib}
\usepackage{color}
\usepackage{xspace}
\definecolor{white}{rgb}{1.00,1.00,1.00}

\shorttitle{Environments of gravitational lenses}
\shortauthors{Moustakas et~al.}

\newenvironment{inlinefigure}{
\def\@captype{figure}
\noindent\begin{minipage}{0.999\linewidth}\begin{center}}
{\end{center}\end{minipage}\smallskip}

\begin{document}
\bibliographystyle{apj}

\slugcomment{Submitted to the AEGIS ApJL Special Issue}

\def\ltsima{$\; \buildrel < \over \sim \;$}
\def\lsim{\lower.5ex\hbox{\ltsima}}
\def\gtsima{$\; \buildrel > \over \sim \;$}
\def\gsim{\lower.5ex\hbox{\gtsima}}

\def\tabcolsep{\fill}

\def\egsarea{$650$\,arcmin$^2$}
\def\Vband{V_{\rm 606}}
\def\Iband{I_{\rm 814}}

\def\CrossIAUname{HST\,J141735+52264} 
\def\DewdropIAUname{HST\,J141820+52361}
\def\AnchorIAUname{HST\,J141833+52435}
\def\FlourishIAUname{HST\,J141807+52302}
\def\QuotesIAUname{HST\,J142052+53065}
\def\DotsIAUname{HST\,J141759+52351}
\def\ColonIAUname{HST\,J142053+53060}

\title{A Strong-Lens Survey in AEGIS: the influence of large scale
structure} 

\author{Leonidas A.~Moustakas\altaffilmark{1}}
\author{Phil Marshall\altaffilmark{2}}
\author{Jeffrey A.~Newman\altaffilmark{3}$^,$\altaffilmark{11}}

\author{Alison L.~Coil\altaffilmark{4}$^,$\altaffilmark{12}}
\author{Michael C.~Cooper\altaffilmark{5}}
\author{Marc Davis\altaffilmark{5}}
\author{Christopher D.~Fassnacht\altaffilmark{6}}
\author{Puragra Guhathakurta\altaffilmark{8}}
\author{Andrew Hopkins\altaffilmark{9}}
\author{Anton Koekemoer\altaffilmark{10}}
\author{Nicholas P.~Konidaris\altaffilmark{8}}
\author{Jennifer M.~Lotz\altaffilmark{11}$^,$\altaffilmark{13}}
\author{Christopher N.~A.~Willmer\altaffilmark{4}}

\altaffiltext{1}{JPL/Caltech, 4800 Oak Grove Dr, MS\,169-327, Pasadena, CA 91109  {\tt leonidas@jpl.nasa.gov}}
\altaffiltext{2}{KIPAC, P.O. Box 20450, MS29, Stanford, CA 94309 {\tt pjm@slac.stanford.edu}} 
\altaffiltext{3}{INPA, LBNL, Berkeley, CA 94720{\tt janewman@lbl.gov}}
\altaffiltext{4}{S.O., University of Arizona, Tucson, AZ 85721 {\tt
acoil, cnaw@as.arizona.edu}}
\altaffiltext{5}{Department of Astronomy, U.C.~Berkeley, Berkeley, CA
94720 {\tt cooper, marc@astron.berkeley.edu}}
\altaffiltext{6}{Department of Physics, U.C.~Davis, Davis, CA 95616 {\tt fassnacht@physics.ucdavis.edu}}
\altaffiltext{8}{U.C.O./Lick Observatory, UCSC, Santa Cruz, CA 95064
{\tt raja, npk@ucolick.org}}
\altaffiltext{9}{School of Physics, University of Sydney NSW,
Australia {\tt ahopkins@anu.phyast.pitt.edu}}
\altaffiltext{10}{Space Telescope Science Institute, Baltimore, MD
21218 {\tt koekemoe@stsci.edu}}
\altaffiltext{11}{NOAO, 950 North Cherry Street, Tucson, AZ 85719 {\tt lotz@noao.edu}}
\altaffiltext{12}{{\em Hubble} Fellow}
\altaffiltext{13}{Goldberg Fellow}

\begin{abstract}
  We report on the results of a visual search for galaxy-scale strong
  gravitational lenses over \egsarea{} of \emph{HST}/ACS imaging in the
  Extended Groth Strip (EGS).  These deep F606W- and F814W-band
  observations are in the DEEP2-EGS field.
  In addition to a previously-known Einstein Cross also found by our
  search (the ``Cross,'' \CrossIAUname{}, with $z_{\rm lens}=0.8106$ and
  a published $z_{\rm source}=3.40$), we identify two new strong
  galaxy-galaxy lenses
  with multiple extended arcs.  The first, \DewdropIAUname{} (the
  ``Dewdrop''; $z_{\rm lens}=0.5798$),
  lenses two distinct extended sources into two pairs of arcs ($z_{\rm
  source}=0.9818$ by nebular [O\,{\scriptsize II}] emission),
  while the second, \AnchorIAUname{} (the ``Anchor'';
  $z_{\rm lens}=0.4625$), produces a single pair of arcs (source
  redshift not yet known).
  Four less convincing arc/counter-arc and two-image lens candidates
  are also found and presented for completeness.  All three definite
  lenses are fit reasonably well by simple singular isothermal 
  ellipsoid models including external 
  shear, giving $\chi^2_{\nu}$ values close to unity.    
  Using the three-dimensional line-of-sight (LOS)
  information on galaxies from the DEEP2 data, we calculate the
  convergence and shear contributions $\kappa_{\rm los}$ and
  $\gamma_{\rm los}$ to each lens, assuming singular isothermal sphere
  halos truncated at 200\,$h^{-1}$\,kpc.  These are compared against a
  robust measure of local environment, $\delta_{3}$, a normalized
  density that uses the distance to the third nearest neighbor.  We
  find that even strong lenses in demonstrably underdense local
  environments may be considerably affected by LOS contributions,
  which in turn, under the adopted assumptions, may be underestimates
  of the effect of large scale structure.  
\end{abstract}



\section{Introduction}\label{sec:intro}

Galaxy-scale gravitational lenses have many astrophysical and
cosmological applications.  These rely on the ability to construct
robust and accurate gravitational lens models.  However, the
contribution of the large-scale structure along the line of
sight (LOS) between the observer and the source is often unknown,
though it may be significant.  In particular, though lens models may
detect the influence of the distorting effects of environmental shear
($\gamma$) in a preferred direction, models of even the most
richly-constrained Einstein Rings with \emph{Hubble Space Telescope}
images \citep[e.g.][]{dye:05,wayth:05,koopmans:06} are still subject
to the mass-sheet degeneracy due to extra field convergence
($\kappa$), which can lead to incorrect lens masses
\citep[e.g.][]{kochanek:04}.  Indeed, lens galaxies are often massive
early-type galaxies, which are generally found in groups or clusters.
The most famous example is the two-image lensed QSO\,Q0957+561
\citep{walsh:79,young:80}.  The determination of $H_0$ from this
system depends crucially on correctly modeling the galaxy cluster
surrounding the primary lensing galaxy \citep[e.g.][]{keeton:00}.
Several other lens-galaxy groups and environments have been studied in
detail \citep{kundic:97a, kundic:97b, tonry:98, tonry:99,
  fassnacht:02, morgan:05, williams:05, momcheva:06, fassnacht:06,
  auger:06}, with sometimes inconclusive results.  In analyses such as
in \citet{keeton:04}, through mock lens realizations, it is shown how
local environment may affect key applications of lenses.  They argue
that $H_0$ and $\Omega_{\Lambda}$ may be overestimated, the expected
ratio of four-image to two-image lenses may be underestimated, and
predictions for millilensing by dark matter substructure may be off by
significant amounts.  Other theoretical work \citep{barkana:96,
  metcalf:05, wambsganss:05} suggests that \emph{all} matter along a
line of sight can be important.

In the emergent era of large-solid angle, densely-sampled
spectroscopic surveys that may include strong lenses, both
environmental and large scale structure effects can be explored
quantitatively.  The DEIMOS spectroscopy of the Extended Groth Strip
(EGS) is particularly well-suited to this task, and is employed here
to both discover new strong galaxy-lenses, and to begin addressing the
quantitative effect of environment in their behavior.

The DEEP2-EGS field is a 120$\times$30\,arcmin strip, the focus of the
``All-wavelength EGS International Survey'' (AEGIS), includes deep
CFHT $BRI$ imaging \citep{coil:04b} and Keck/DEIMOS spectroscopy of
nearly 14\,000 galaxies to date. The spectroscopy is $\sim75\%$
complete to $R_{\rm AB}<24.1$.  For the analysis here, we only employ
the most certain redshift assignments \citep{coil:04a}.  Deep
\emph{HST}/ACS imaging of nearly \egsarea{} over 63 stitched tiles reach
$\Vband=28.75$ and $\Iband=28.10$ (AB, 5$\sigma$ point source; Davis
et al, this issue).
These data lend themselves to two different techniques for searching
for heretofore-unknown gravitational lenses: spectroscopically and
visually.  The spectroscopic redshifts are supplemented as necessary
with photometric redshifts measured from deep KPNO $UBVRI$ imaging
(A.~Hopkins et al., in prep).  

The spectroscopic approach of searching for ``anomalous'' emission
lines in early-type spectra has some history
\citep[e.g.][]{warren:96}, and has recently proved to be
spectacularly successful when applied to SDSS spectroscopy
\citep{bolton:04, willis:05} with \emph{HST}/ACS followup
\citep{bolton:05, bolton:06, treu:06}.  Explicitly spectroscopic searches for
lenses in the DEEP2 data will be explored elsewhere.

In the imaging domain, one may hope to search for lens candidates by
some automated algorithm, or by visual inspection
\citep[e.g.][]{ratnatunga:95, zepf:97, fassnacht:04}.  The more
quantitative and objective automated approach may eventually be
preferred (especially for datasets larger than the one considered
here), but would, however, require a training set.  The EGS ACS data
described here is used for just this purpose in a separate work
(Marshall et al.~in prep) as a precursor to searching the entire
\emph{HST} imaging
dataset.\footnote{{\tt{http://www.slac.stanford.edu/\~{}pjm/HAGGLeS}}}
Towards that goal we have undertaken a search for lenses by
purely visual inspection.  

The lens-search methodology is described in \S\,\ref{sec:data}.  The
newly discovered lenses and the modeling results are given in
\S\,\ref{sec:lensmod}, while measurements of the local and LOS
environments of the lenses are given in \S\,\ref{sec:lensenv}.
Discussion and conclusions are the subject of \S\,\ref{sec:discuss}.
A concordance flat cosmology with $\Omega_{\Lambda}=1-\Omega_{\rm
  m}=0.7$ and $H_0=100\,h$\,km\,s$^{-1}$\,Mpc$^{-1}$ with $h=0.7$ is
used throughout.  Unless otherwise stated, all magnitudes are in the
AB system.


\section{Lens-search methodology}\label{sec:data}

The search for gravitational lens candidates was conducted by-eye.
Three-color images of all of the ACS tiles were built following the
\citet{lupton:04} algorithm, using the photometric zeropoints to
provide the relative scale factors, and using the mean of the F606W
and F814W images for the green channel.  The full ACS dataset was
inspected repeatedly in the color images at full resolution, with
plausible candidates classified with grades of ``A'' or ``B'' and
marked for further inspection.
Object coordinates were then matched against the DEEP2 spectroscopic
catalog, which includes a ``serendipitous feature'' flag, for possible
anomalous, higher-redshift emission lines.  Emission from a source
behind the Dewdrop lens (described below) was found in this way.

\begin{inlinefigure}
  \begin{center}
    \resizebox{\textwidth}{!}{\includegraphics{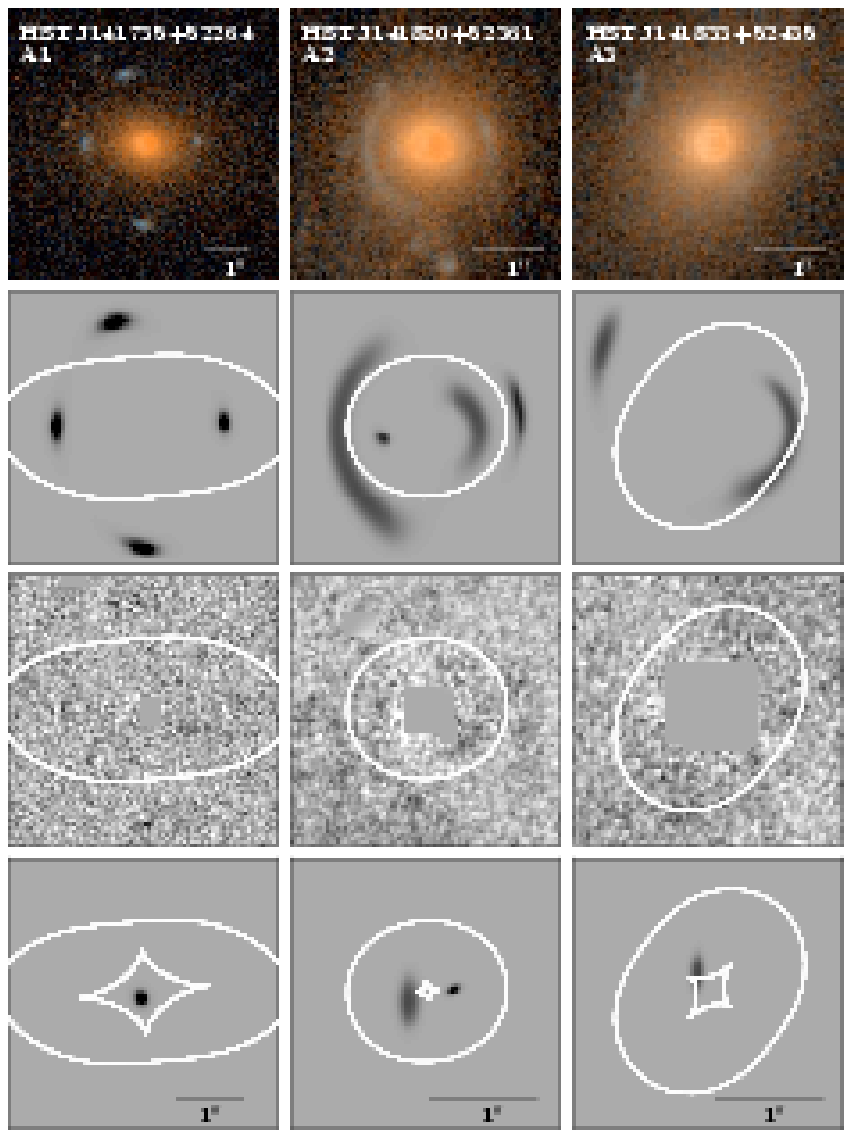}}
    \figcaption{\label{fig:lenses}The three most plausible lenses from
      this survey, from left to right the Cross, the Dewdrop, and the
      Anchor (see text).  From top to bottom, we show the discovery
      image, the lensed image model, the residuals by subtraction with
      the (lens-galaxy-removed) imaging data, and the reconstructed
      source.  All panels, including the source-plane one, are
      approximately 3\,arcsec on a side, with the exact dimensions shown
      by the scale bars.  The image-plane critical curves and the
      source-plane caustics are shown in the third and fourth rows,
      respectively.}
  \end{center}
\end{inlinefigure}

\begin{inlinefigure}
  \begin{center}
    \resizebox{\textwidth}{!}{\includegraphics{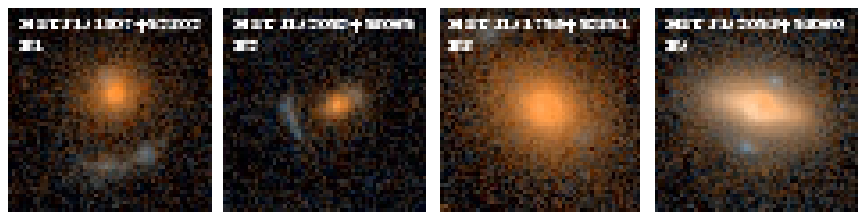}}
    \figcaption{\label{fig:otherlenses} Additional lens candidates based on
      visual inspection.  These are not yet bolstered by spectroscopy, but
      will be targetted when possible.  The left two are candidate
      arc/counter-arc lenses, whereas the right two are candidate
      two-image lenses. Images are 3 arcsec square.}
  \end{center} 
\end{inlinefigure}

\section{Lenses \& Models}\label{sec:lensmod}

In addition to a previously known Einstein Cross, we find two new
unambiguous strong galaxy-galaxy lenses (Fig.~\ref{fig:lenses}).  Four
additional plausible lens candidates (Fig.~\ref{fig:otherlenses}) are
also reported on. 
Here we describe the lens modeling and the model results for each
lens.

\subsection{Lens modeling and source reconstruction}

The lensed sources in the EGS all appear to be blue and extended, and
are likely star forming galaxies at high redshift ($z \sim 1$). We
therefore take the image pixels as our data (rather than simply 
image-centroid positions), and predict the image using a simple ray
tracing forward from the source plane, followed by a PSF convolution.
We first subtract the lens galaxy light using a tilted 2D Moffat
profile,\footnote{The Moffat function is a modified Lorentzian with
  variable power law index.  The fit is done with the {\tt MPFIT} IDL
  suite of C.~Markwardt.} and mask the very center of the lens galaxy
where some residual flux remains.  It is important that the unmasked
region contain not only the lensed images but also the clean pixels
that do \emph{not} have lensed features.  These clean pixels contain
at least as much information as the ones with lensed flux, vetoing
models that predict images where there are none.  For the projected
mass profile of the lens we adopt a singular isothermal ellipsoid
\citep[SIE;][]{kormann:94} model, plus an external shear component.
Using a Markov chain Monte Carlo procedure presented in detail
elsewhere (Marshall et~al.~in prep), the position, ellipticity,
orientation and mass of the lens, external shear amplitude and the
direction, position, ellipticity, orientation and Sersic profile
parameters of the source are all fit to the data.
Since we are interested in accurate estimation of the lens
environment, we apply a prior on the orientation of the lens
ellipticity to reflect the expected correlation with the lens light
\citep[e.g.][]{koopmans:06}.  

\input{tab1.tex}

\subsection{\CrossIAUname{} (A1 -- Cross)}

This lens was originally discovered by \citet{ratnatunga:95} by visual
inspection of the \emph{HST}/WFPC2 Medium Deep Survey (MDS) data.  The
lens redshift is $z_{\rm lens}=0.8106$ (Table~\ref{tab:lenses}), and
the source is at $z_{\rm source}=3.4$ \citep{crampton:96}.  The large
Einstein radius $\theta_{\rm E}=1.447$\,arcsec and the four-image
configuration require a large enclosed mass and a significant amount
of external shear, $\gamma_{\rm mod}=0.080$, a result consistent with
\citet{treu:04}.  The best-fit model shows very small residuals at the
two outer images, a feature corrected for by \citet{treu:04} with a
potential gradient that is presumably associated with a nearby
structure. The mass and external shear are not affected by this
correction.

\subsection{\DewdropIAUname{} (A2 -- Dewdrop)}

The Dewdrop lens at $z_{\rm lens}=0.5798$ lenses two distinct sources
into two pairs of arcs.  The Keck/DEIMOS spectrum of the system
reveals anomalous [O\,{\scriptsize II}] nebular emission at $z_{\rm
source}=0.9818$ (Fig.~\ref{fig:dewspec}).  The sources in the Dewdrop
system are part of a remarkable irregular and loose association of star
formation knots and diffuse emitting material that extends over more
than 10\,arcsec, or more than 80\,kpc comoving in size.  

\subsection{\AnchorIAUname{} (A3 -- Anchor)}

The Anchor system exhibits a pair of arcs created by a lens at a
redshift of $z_{\rm lens}=0.4625$.  The best-fitting lens model
requires a significant external shear contribution (see
Table~\ref{tab:lenses}), as might be expected from the position and
shape of the counter-image to the main arc.

\begin{inlinefigure}
  \begin{center}
    \resizebox{\textwidth}{!}{\includegraphics{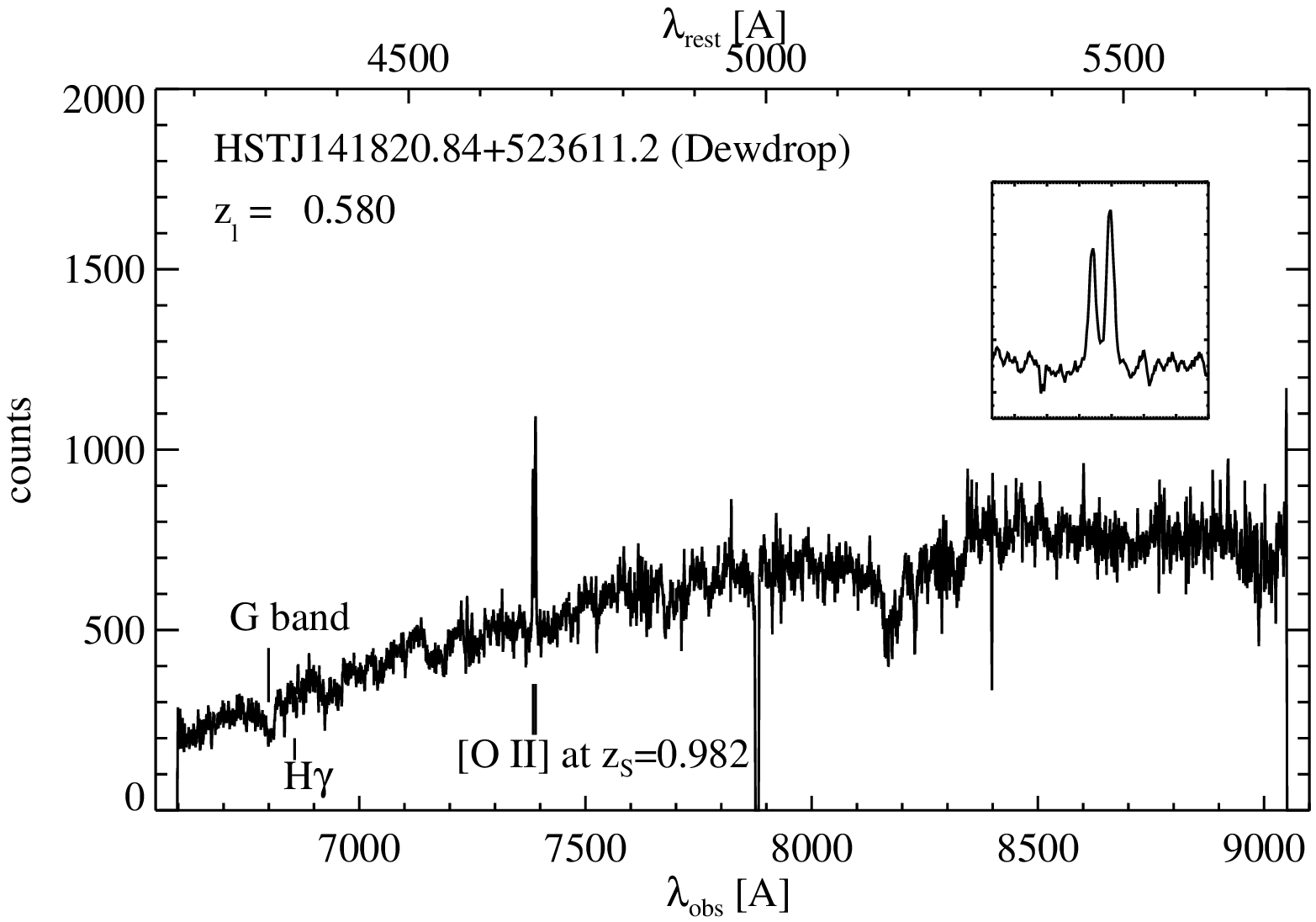}}
    \figcaption{{\label{fig:dewspec}The DEIMOS spectrum of the Dewdrop
        lens clearly shows an ``anomalous'' doublet emission line
        (insert), which is readily identified as [O\,{\scriptsize II}]
        at $z_{\rm source}=0.9818$.}}
  \end{center}
\end{inlinefigure}

\subsection{Additional lens candidates}

In Fig.~\ref{fig:otherlenses} and Table~\ref{tab:lenses} we identify
four additional visually-identified lens candidates.  Only two of the
four presently have redshifts measured, and require further
spectroscopic followup.  These are presented for completeness, and do
not affect the scope or results of this paper.

\section{The Environments of the Lenses}\label{sec:lensenv}

We explore the environments of the lenses in two different ways.  The
first makes use of a relatively unbiased measure of the very local
environment of any one galaxy, dubbed $\delta_{3}$ and explored in
detail in \citet{cooper:05, cooper:06}.  
This parameter is derived from the distance to the third-nearest
neighbor among the galaxies within 1000\,km\,s$^{-1}$ along the line
of sight, and scales as the inverse of the cube of this distance.
Thus, more concentrated environments have larger values of $\delta_3$.
The typical uncertainties in individual measures of
$\delta_{3}$ are $\sim0.5$\,dex.  We only compute this measure for
galaxies with spectroscopic redshifts.  

\begin{inlinefigure}
  \begin{center}
    \resizebox{\textwidth}{!}{\includegraphics{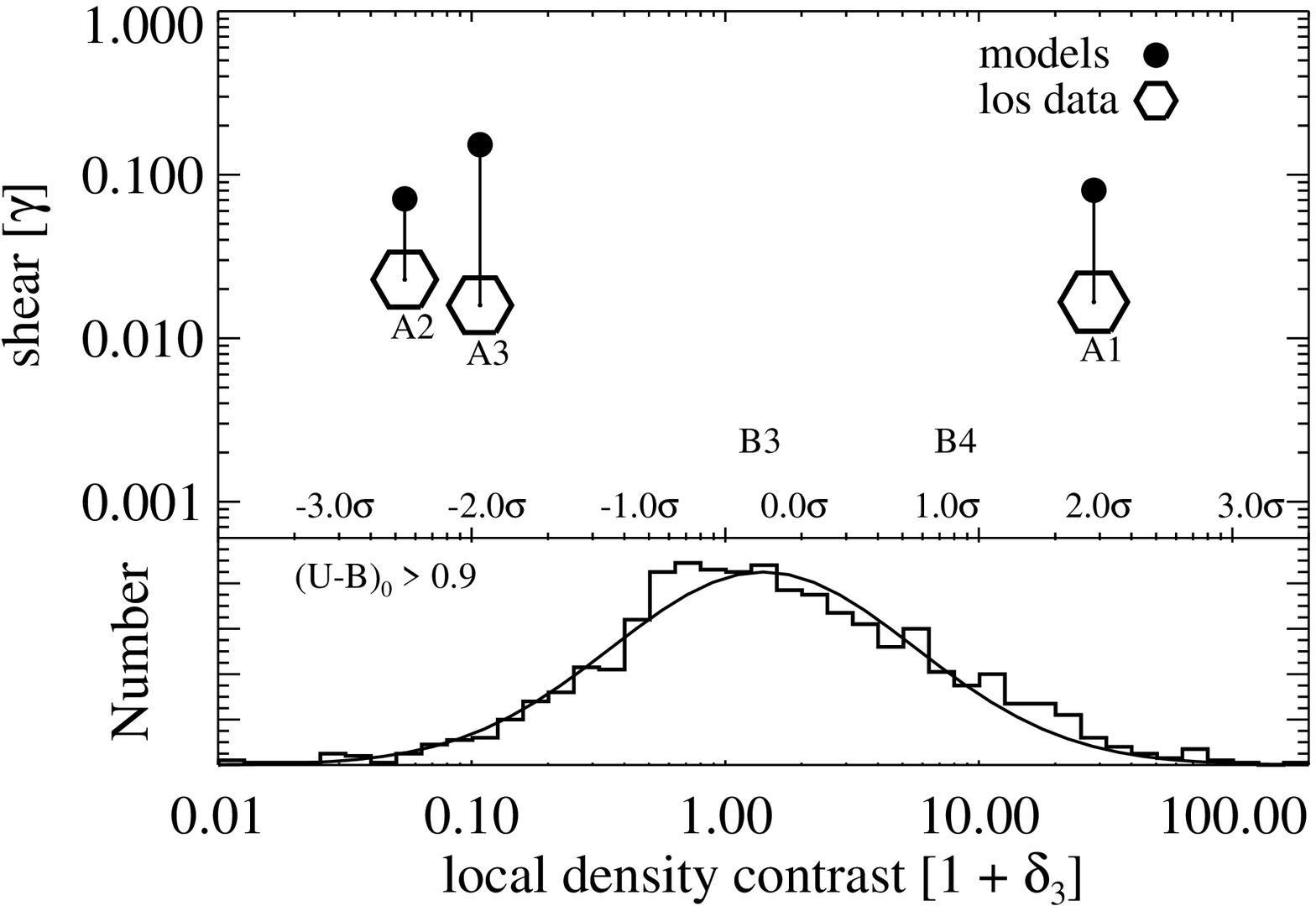}}
    \figcaption{{\label{fig:drho} The lower panel shows the
        distribution of the local-enviromental measure 1+$\delta_{3}$
        \citep{cooper:05}, for ``red sequence'' galaxies with
        rest-frame colors $(U-B)_{0}>0.9$.  (All lenses are found to
        satisfy the same color-criterion).  Based on a gaussian fit to
        the distribution, $N\sigma$ positions are marked in the x-axis
        line above, as a guide.  The upper panel shows the lens-model
        and line-of-sight shear for each object, as the filled-circles
        and open hexagons, respectively.  The size of the hexagons
        corresponds to the calculated line of sight convergence
        $\kappa_{\rm los}$.  The 1+$\delta_{3}$ values of the
        lens-candidates B3 and B4 are shown as well.}}
  \end{center}
\end{inlinefigure}

As a second probe of lens environment we model the contribution to
the lensing potential due to individual neighboring galaxies using
simple analytic mass distributions.  
We calculate the convergence $\kappa_{\rm los}$ and shear $\gamma_{\rm
  los}$ line-of-sight contribution by all galaxies within a projected
separation of 200\,$h^{-1}$\,kpc from the lens galaxies, out to the
redshift of the source.  We treat each galaxy as an isolated halo,
undoubtedly neglecting the effect of group halos and other structures.
Assuming that we can approximate each galaxy $i$ as a singular
isothermal sphere (SIS), we have $\kappa_i = b_i / 2 r_i$,
where $r$ is the projected distance from the lens and 
$b$ is the ``lens strength'' for a background source at angular
diameter distances of $D_{\rm s}$ from the observer and $D_{\rm ls}$
from the lens, 
\begin{equation}
b = 4\pi\left( {{\sigma_{\rm dm}}\over{c}} \right)^2
{{D_{\rm ls}}\over{D_{\rm s}}}. 
\end{equation}
The central dark matter velocity dispersion $\sigma_{\rm dm}$ of each
galaxy is assumed to be the same as the central stellar velocity
dispersion, which is derived from the estimated rest-frame $B$-band
(Vega) magnitude of each galaxy using the Faber-Jackson relationship
as given in \citet{mitchell:05} (see also \citealt{jonsson:06}). (We
neglect the dispersion in this relation).
The total shear contribution is the ``headless-vector'' sum of the
shears, $\bar{\gamma}_{\rm los}=\Sigma\bar{\gamma_{i}}$, while the
total convergence is a scalar sum: $\kappa_{\rm
  los}=\Sigma\kappa_{i}$.
It is worth noting that if at large radii the profiles are steeper
than SIS (such as NFW), the convergence contribution will be smaller
overall than the shear.
These measurements
are given in Table~\ref{tab:lenses} and discussed in the last section.


\section{Discussion \& Conclusions}\label{sec:discuss}

The numbers of definite lenses reported on here is consistent with
other surveys.  For example, \citet{bolton:06} find that $\sim$0.1\%
of luminous red galaxies are very likely to be strong galaxy-galaxy
lenses, although special lines of sight can have much higher lensing
rates \citep[e.g.][]{fassnacht:06b}.
The rate
above then suggests that there should be $\sim$4 strong lenses in this
survey, which is a good match to our three.  

The main conclusions of this work can be drawn by an examination of
Fig.~\ref{fig:drho}.  The Cross is in a fairly overdense local
environment, which is consistent with this lens being associated with
the $z\approx0.8$ sheet described in \citet{koo:96} and \citet{im:02}.
Given this, the shear of $\sim$10\% required by the model (see also
\citealt{treu:04}) seems plausible.  What seems surprising is that
even though both the Dewdrop and Anchor lenses are in
\emph{under}-dense environments locally, they still require relatively
large shear contributions to produce good fits.  In all three cases,
we also note the large discrepancy between the modeled and
LOS-predicted shear values.  To explore this, we ran lens models with
external shear and orientation \emph{restricted to the predicted
  values}, and then examining the resulting models, and particularly
the fit $\chi_{\nu}^2$.  All three new models require lenses with much
higher ellipticity than the light suggests, though in the Dewdrop and
the Anchor the formal $\chi_{\nu}^2$ remains plausible given the
constraints, $\chi_{\nu}^2=1.02$ and $1.04$ (or underfit by $\sim$1-
and $\sim$2-$\sigma$), respectively.  The new Cross fit, however, is
strongly ruled out with $\chi_{\nu}^2=2.00$ (or by $\sim$75-$\sigma$).
This suggests that at least in this case, the inferred LOS influence
by SIS dark matter halos is insufficient, and that the large scale
structure ``sheet'' must have an important additional effect.

Our conclusions may be summarized as follows: 
1. We have discovered
two new strong galaxy-galaxy lenses by visual inspection, with
reasonable lens models and source reconstructions.
2. These lenses are drawn from a range of local-density
environments, which do not necessarily reflect the influence of
unassociated large scale structure.
3. In at least the case of the Cross, the known large scale structure
sheet at the redshift of the lens, which is not formally accounted for
in the LOS calculation, has a demonstrable effect on the lens model.


\acknowledgments

We thank Maru\v{s}a Brada\v{c} for discussions.  LAM thanks Russell
Mirabelli for expert assistance with a GIMP script facilitating the
inspection of the ACS data, and UC Berkeley and UC Santa Cruz for
their frequent hospitality during the course of this work.  The work
of LAM was carried out at Jet Propulsion Laboratory, California
Institute of Technology, under a contract with NASA.  The work of PJM
was supported in part by the U.S. Department of Energy under contract
number DE-AC02-76SF00515.  JAN and ALC are supported by NASA through
the \emph{Hubble} Fellowship grants HF-011065.01-A and HF-01182.01-A,
respectively.

\end{document}

%% file: tab1.tex
\begin{table*}[t]
\begin{center}
\scriptsize
\caption{\label{tab:lenses} EGS lenses: data, environment, \& models}
\begin{tabular*}{\linewidth}{@{\extracolsep{\fill}}cc cc l cc cc  c ccccr c ccccc}
\hline\hline
\multicolumn{9}{c}{Data}       & \multicolumn{1}{c}{~~~} & 
\multicolumn{5}{c}{Environment} & \multicolumn{1}{c}{~~~} & 
\multicolumn{5}{c}{Models} \\
\cline{ 1- 9}
\cline{11-15}
\cline{17-21} 
\multicolumn{1}{c}{ID} &
\multicolumn{1}{c}{Alias} &
\multicolumn{1}{c}{RA} &
\multicolumn{1}{c}{Dec} &
\multicolumn{1}{c}{$z_{\rm lens}$} &
\multicolumn{1}{c}{$R$} &
\multicolumn{1}{c}{$M_{\rm B}^{c}$} &
\multicolumn{1}{c}{} &
\multicolumn{1}{c}{$z_{\rm source}$} &
\multicolumn{1}{c}{} & 
\multicolumn{1}{c}{$\log(1+\delta_{3})$} & 
\multicolumn{1}{c}{N$_{\rm los}$} & 
\multicolumn{1}{c}{$\kappa_{\rm los}$} &
\multicolumn{1}{c}{$\gamma_{\rm los}$} &
\multicolumn{1}{c}{$\theta_{\gamma_{\rm los}}$} &
\multicolumn{1}{c}{} & 
\multicolumn{1}{c}{$\theta_{\rm E}$} &
\multicolumn{1}{c}{$\sigma_{\rm SIS}$} &
\multicolumn{1}{c}{$\gamma_{\rm mod}$} &
\multicolumn{1}{c}{$\theta_{\gamma_{\rm mod}}$} &
\multicolumn{1}{c}{$\chi^2_{\nu}$} \\
\multicolumn{1}{c}{} &
\multicolumn{1}{c}{} &
\multicolumn{1}{c}{(J2000)} &
\multicolumn{1}{c}{(J2000)} &
\multicolumn{1}{c}{} &
\multicolumn{1}{c}{(AB)} &
\multicolumn{1}{c}{(AB)} &
\multicolumn{1}{c}{} & 
\multicolumn{1}{c}{} &
\multicolumn{1}{c}{} & 
\multicolumn{1}{c}{} & 
\multicolumn{1}{c}{} & 
\multicolumn{1}{c}{} &
\multicolumn{1}{c}{} &
\multicolumn{1}{c}{($^{\circ}$E)} &
\multicolumn{1}{c}{} & 
\multicolumn{1}{c}{$('')$} &
\multicolumn{1}{c}{km\,s$^{-1}$} &
\multicolumn{1}{c}{} &
\multicolumn{1}{c}{($^{\circ}$E)} &
\multicolumn{1}{c}{} 
\\
\cline{ 1- 9}
\cline{11-15}
\cline{17-21} 
A1 & Cross    &  14:17:35.72 & 52:26:46.3& $0.8106$ & $21.38$ & $-21.25$   & & $3.40  $ & & $+1.453$ & 36 & 0.17 & 0.02 & $  78$ & & 1.45 & 292.8 & 0.080 & $115.3$ & 1.081\\
A2 & Dewdrop  &  14:18:20.77 & 52:36:11.3& $0.5798$ & $20.55$ & $-20.35$   & & $0.9818$ & & $-1.260$ & 46 & 0.10 & 0.02 & $ 140$ & & 0.67 & 260.6 & 0.071 & $101.4$ & 1.005\\
A3 & Anchor   &  14:18:33.11 & 52:43:52.6& $0.4625$ & $20.45$ & $-19.47$   & & $  ... $ & & $-0.960$ & 52 & 0.09 & 0.02 & $ 146$ & & 0.83 & 248.9 & 0.153 & $140.8$ & 0.933\\
B1 & Flourish &  14:18:07.32 & 52:30:29.8& $0.847^a$ & $22.58$ & $(-17.8)$ & & $  ... $ & & $ ... $  & ...& ... & ... & ...  & & $...$&$...$&$...$&$...$&$...$\\
B2 & Quotes   &  14:20:52.01 & 53:06:57.2& $0.601^b$ & $23.82$ & $(-16.8)$ & & $  ... $ & & $ ... $  & ...& ... & ... & ...  & & $...$&$...$&$...$&$...$&$...$\\
B3 & Dots     &  14:17:59.01 & 52:35:14.8& $0.6863$ & $21.50$ & $-20.24$   & & $  ... $ & & $+0.109$ & ...& ... & ... & ...  & & $...$&$...$&$...$&$...$&$...$\\
B4 & Colon    &  14:20:53.89 & 53:06:07.0& $0.3545$ & $20.61$ & $-18.47$   & & $  ... $ & & $+0.880$ & ...& ... & ... & ...  & & $...$&$...$&$...$&$...$&$...$\\
\end{tabular*}
\end{center}
\smallskip
\scriptsize
$a$: $\sigma_z=0.067$ \& $b$: $\sigma_z=0.24$ (A.~Hopkins et al., in
prep); $c$: parenthetical quantities are based on photometric redshifts.\\
\end{table*}

%% file: ms.bbl
\begin{thebibliography}{43}
\expandafter\ifx\csname natexlab\endcsname\relax\def\natexlab#1{#1}\fi

\bibitem[{{Auger} {et~al.}(2006){Auger}, {Fassnacht}, {Abrahamse}, {Lubin}, \&
  {Squires}}]{auger:06}
{Auger}, M.~W., {Fassnacht}, C.~D., {Abrahamse}, A.~L., {Lubin}, L.~M., \&
  {Squires}, G.~K. 2006, astro-ph/0603448

\bibitem[{{Bar-Kana}(1996)}]{barkana:96}
{Bar-Kana}, R. 1996, ApJ, 468, 17

\bibitem[{{Bolton} {et~al.}(2005){Bolton}, {Burles}, {Koopmans}, {Treu}, \&
  {Moustakas}}]{bolton:05}
{Bolton}, A.~S., {Burles}, S., {Koopmans}, L.~V.~E., {Treu}, T., \&
  {Moustakas}, L.~A. 2005, ApJL, 624, 21

\bibitem[{{Bolton} {et~al.}(2006){Bolton}, {Burles}, {Koopmans}, {Treu}, \&
  {Moustakas}}]{bolton:06}
---. 2006, ApJ, 0, 0

\bibitem[{{Bolton} {et~al.}(2004){Bolton}, {Burles}, {Schlegel}, {Eisenstein},
  \& {Brinkmann}}]{bolton:04}
{Bolton}, A.~S., {Burles}, S., {Schlegel}, D.~J., {Eisenstein}, D.~J., \&
  {Brinkmann}, J. 2004, AJ, 127, 1860

\bibitem[{{Coil} {et~al.}(2004{\natexlab{a}}){Coil}, {Newman}, {Kaiser},
  {Davis}, {Ma}, {Kocevski}, \& {Koo}}]{coil:04b}
{Coil}, A.~L., {Newman}, J.~A., {Kaiser}, N., {Davis}, M., {Ma}, C.-P.,
  {Kocevski}, D.~D., \& {Koo}, D.~C. 2004{\natexlab{a}}, ApJ, 617, 765

\bibitem[{{Coil} {et~al.}(2004{\natexlab{b}})}]{coil:04a}
{Coil}, A.~L., {et~al.} 2004{\natexlab{b}}, ApJ, 609, 525

\bibitem[{{Cooper} {et~al.}(2005{\natexlab{a}}){Cooper}, {Newman}, {Madgwick},
  {Gerke}, {Yan}, \& {Davis}}]{cooper:05}
{Cooper}, M.~C., {Newman}, J.~A., {Madgwick}, D.~S., {Gerke}, B.~F., {Yan}, R.,
  \& {Davis}, M. 2005{\natexlab{a}}, ApJ, 634, 833

\bibitem[{{Cooper} {et~al.}(2005{\natexlab{b}})}]{cooper:06}
{Cooper}, M.~C., {et~al.} 2005{\natexlab{b}}, ArXiv Astrophysics e-prints

\bibitem[{{Crampton} {et~al.}(1996){Crampton}, {Le Fevre}, {Hammer}, \&
  {Lilly}}]{crampton:96}
{Crampton}, D., {Le Fevre}, O., {Hammer}, F., \& {Lilly}, S.~J. 1996, A\&A,
  307, L53

\bibitem[{{Dye} \& {Warren}(2005)}]{dye:05}
{Dye}, S., \& {Warren}, S.~J. 2005, ApJ, 623, 31

\bibitem[{{Fassnacht} {et~al.}(2006{\natexlab{a}}){Fassnacht}, {Gal}, {Lubin},
  {McKean}, {Squires}, \& {Readhead}}]{fassnacht:06}
{Fassnacht}, C.~D., {Gal}, R.~R., {Lubin}, L.~M., {McKean}, J.~P., {Squires},
  G.~K., \& {Readhead}, A.~C.~S. 2006{\natexlab{a}}, \apj, 642, 30

\bibitem[{{Fassnacht} \& {Lubin}(2002)}]{fassnacht:02}
{Fassnacht}, C.~D., \& {Lubin}, L.~M. 2002, \aj, 123, 627

\bibitem[{{Fassnacht} {et~al.}(2004){Fassnacht}, {Moustakas}, {Casertano},
  {Ferguson}, {Lucas}, \& {Park}}]{fassnacht:04}
{Fassnacht}, C.~D., {Moustakas}, L.~A., {Casertano}, S., {Ferguson}, H.~C.,
  {Lucas}, R.~A., \& {Park}, Y. 2004, ApJL, 600, L155

\bibitem[{{Fassnacht} {et~al.}(2006{\natexlab{b}})}]{fassnacht:06b}
{Fassnacht}, C.~D., {et~al.} 2006{\natexlab{b}}, ArXiv Astrophysics e-prints

\bibitem[{{Im} {et~al.}(2002)}]{im:02}
{Im}, M., {et~al.} 2002, ApJ, 571, 136

\bibitem[{{J{\"o}nsson} {et~al.}(2006){J{\"o}nsson}, {Dahl{\'e}n}, {Goobar},
  {Gunnarsson}, {M{\"o}rtsell}, \& {Lee}}]{jonsson:06}
{J{\"o}nsson}, J., {Dahl{\'e}n}, T., {Goobar}, A., {Gunnarsson}, C.,
  {M{\"o}rtsell}, E., \& {Lee}, K. 2006, ApJ, 639, 991

\bibitem[{{Keeton} {et~al.}(2000){Keeton}, {Falco}, {Impey}, {Kochanek},
  {Leh{\'a}r}, {McLeod}, {Rix}, {Mu{\~n}oz}, \& {Peng}}]{keeton:00}
{Keeton}, C.~R., {Falco}, E.~E., {Impey}, C.~D., {Kochanek}, C.~S.,
  {Leh{\'a}r}, J., {McLeod}, B.~A., {Rix}, H.-W., {Mu{\~n}oz}, J.~A., \&
  {Peng}, C.~Y. 2000, \apj, 542, 74

\bibitem[{{Keeton} \& {Zabludoff}(2004)}]{keeton:04}
{Keeton}, C.~R., \& {Zabludoff}, A.~I. 2004, ApJ, 612, 660

\bibitem[{{Kochanek}(2004)}]{kochanek:04}
{Kochanek}, C.~S. 2004, astro-ph/0407232

\bibitem[{{Koo} {et~al.}(1996)}]{koo:96}
{Koo}, D.~C., {et~al.} 1996, ApJ, 469, 535

\bibitem[{{Koopmans} {et~al.}(2006){Koopmans}, {Treu}, {Bolton}, {Burles}, \&
  {Moustakas}}]{koopmans:06}
{Koopmans}, L.~V.~E., {Treu}, T., {Bolton}, A.~S., {Burles}, S., \&
  {Moustakas}, L.~A. 2006, ApJ, 0, 0

\bibitem[{{Kormann} {et~al.}(1994){Kormann}, {Schneider}, \&
  {Bartelmann}}]{kormann:94}
{Kormann}, R., {Schneider}, P., \& {Bartelmann}, M. 1994, \aap, 284, 285

\bibitem[{{Kundic} {et~al.}(1997{\natexlab{a}}){Kundic}, {Cohen}, {Blandford},
  \& {Lubin}}]{kundic:97a}
{Kundic}, T., {Cohen}, J.~G., {Blandford}, R.~D., \& {Lubin}, L.~M.
  1997{\natexlab{a}}, \aj, 114, 507

\bibitem[{{Kundic} {et~al.}(1997{\natexlab{b}}){Kundic}, {Hogg}, {Blandford},
  {Cohen}, {Lubin}, \& {Larkin}}]{kundic:97b}
{Kundic}, T., {Hogg}, D.~W., {Blandford}, R.~D., {Cohen}, J.~G., {Lubin},
  L.~M., \& {Larkin}, J.~E. 1997{\natexlab{b}}, \aj, 114, 2276

\bibitem[{{Lupton} {et~al.}(2004){Lupton}, {Blanton}, {Fekete}, {Hogg},
  {O'Mullane}, {Szalay}, \& {Wherry}}]{lupton:04}
{Lupton}, R., {Blanton}, M.~R., {Fekete}, G., {Hogg}, D.~W., {O'Mullane}, W.,
  {Szalay}, A., \& {Wherry}, N. 2004, PASP, 116, 133

\bibitem[{{Metcalf}(2005)}]{metcalf:05}
{Metcalf}, R.~B. 2005, ApJ, 629, 673

\bibitem[{{Mitchell} {et~al.}(2005){Mitchell}, {Keeton}, {Frieman}, \&
  {Sheth}}]{mitchell:05}
{Mitchell}, J.~L., {Keeton}, C.~R., {Frieman}, J.~A., \& {Sheth}, R.~K. 2005,
  ApJ, 622, 81

\bibitem[{{Momcheva} {et~al.}(2006){Momcheva}, {Williams}, {Keeton}, \&
  {Zabludoff}}]{momcheva:06}
{Momcheva}, I., {Williams}, K., {Keeton}, C., \& {Zabludoff}, A. 2006, \apj,
  641, 169

\bibitem[{{Morgan} {et~al.}(2005){Morgan}, {Kochanek}, {Pevunova}, \&
  {Schechter}}]{morgan:05}
{Morgan}, N.~D., {Kochanek}, C.~S., {Pevunova}, O., \& {Schechter}, P.~L. 2005,
  AJ, 129, 2531

\bibitem[{{Ratnatunga} {et~al.}(1995){Ratnatunga}, {Ostrander}, {Griffiths}, \&
  {Im}}]{ratnatunga:95}
{Ratnatunga}, K.~U., {Ostrander}, E.~J., {Griffiths}, R.~E., \& {Im}, M. 1995,
  ApJL, 453, 5

\bibitem[{{Tonry}(1998)}]{tonry:98}
{Tonry}, J.~L. 1998, \aj, 115, 1

\bibitem[{{Tonry} \& {Kochanek}(1999)}]{tonry:99}
{Tonry}, J.~L., \& {Kochanek}, C.~S. 1999, \aj, 117, 2034

\bibitem[{{Treu} {et~al.}(2006){Treu}, {Koopmans}, {Bolton}, {Burles}, \&
  {Moustakas}}]{treu:06}
{Treu}, T., {Koopmans}, L.~V., {Bolton}, A.~S., {Burles}, S., \& {Moustakas},
  L.~A. 2006, \apj, 640, 662

\bibitem[{{Treu} \& {Koopmans}(2004)}]{treu:04}
{Treu}, T., \& {Koopmans}, L.~V.~E. 2004, ApJ, 611, 739

\bibitem[{{Walsh} {et~al.}(1979){Walsh}, {Carswell}, \& {Weymann}}]{walsh:79}
{Walsh}, D., {Carswell}, R.~F., \& {Weymann}, R.~J. 1979, Nature, 279, 381

\bibitem[{{Wambsganss} {et~al.}(2005){Wambsganss}, {Bode}, \&
  {Ostriker}}]{wambsganss:05}
{Wambsganss}, J., {Bode}, P., \& {Ostriker}, J.~P. 2005, ApJL, 635, L1

\bibitem[{{Warren} {et~al.}(1996){Warren}, {Hewett}, {Lewis}, {Moller},
  {Iovino}, \& {Shaver}}]{warren:96}
{Warren}, S.~J., {Hewett}, P.~C., {Lewis}, G.~F., {Moller}, P., {Iovino}, A.,
  \& {Shaver}, P.~A. 1996, \mnras, 278, 139

\bibitem[{{Wayth} {et~al.}(2005){Wayth}, {Warren}, {Lewis}, \&
  {Hewett}}]{wayth:05}
{Wayth}, R.~B., {Warren}, S.~J., {Lewis}, G.~F., \& {Hewett}, P.~C. 2005,
  MNRAS, 360, 1333

\bibitem[{{Williams} {et~al.}(2005){Williams}, {Momcheva}, {Keeton},
  {Zabludoff}, \& {Lehar}}]{williams:05}
{Williams}, K.~A., {Momcheva}, I., {Keeton}, C.~R., {Zabludoff}, A.~I., \&
  {Lehar}, J. 2005, astro-ph/0511593

\bibitem[{{Willis} {et~al.}(2005){Willis}, {Hewett}, \& {Warren}}]{willis:05}
{Willis}, J.~P., {Hewett}, P.~C., \& {Warren}, S.~J. 2005, MNRAS, 363, 1369

\bibitem[{{Young} {et~al.}(1980){Young}, {Gunn}, {Oke}, {Westphal}, \&
  {Kristian}}]{young:80}
{Young}, P., {Gunn}, J.~E., {Oke}, J.~B., {Westphal}, J.~A., \& {Kristian}, J.
  1980, \apj, 241, 507

\bibitem[{{Zepf} {et~al.}(1997){Zepf}, {Moustakas}, \& {Davis}}]{zepf:97}
{Zepf}, S.~E., {Moustakas}, L.~A., \& {Davis}, M. 1997, ApJL, 474, L1

\end{thebibliography}
